\documentclass[a4paper]{article}  

\def\comment#1{} 
\def\journalfont{\rm}         
\def\jou#1{{\journalfont #1\ }}
\def\joudef#1#2{\def #1{\jou{\ignorespaces #2}}}

\joudef{\aaa}    { Astron.\ Astrophys.}
\joudef{\aip}    { Adv.\ Phys.}
\joudef{\adm}    { adv.\ math.}
\joudef{\am}     { Ann.\ Math.}
\joudef{\apny}   { Ann.\ Phys.\ (N.Y.)}
\joudef{\apj}    { Astrophys.\ J.}
\joudef{\cjp}    { Can.\ J.\ Phys.}
\joudef{\cmp}    { Commun.\ Math.\ Phys.}
\joudef{\cqg}    { Class.\ Quantum Grav.}
\joudef{\grg}    { Gen.\ Rel.\ Grav.}
\joudef{\ijmpd}  { Int.\ J.\ Mod.\ Phys.\ D}
\joudef{\ijtp}   { Int.\ J.\ Theor.\ Phys.}
\joudef{\invm}   { Invent.\ Math.}
\joudef{\jm}     { J.\ Math.}
\joudef{\jmaa}   { J.\ Math.\ Anal.\ Appl.}
\joudef{\jmp}    { J.\ Math.\ Phys.}
\joudef{\jpa}    { J.\ Phys.\ A}
\joudef{\mnras}  { Mon.\ Not.\ R.\ Ast.\ Soc.}
\joudef{\mpla}   { Mod.\ Phys.\ Lett.\ A} 
\joudef{\nature} { Nature}
\joudef{\nc}     { Nuovo Cim.}
\joudef{\ncb}    { Nuovo Cim. B}
\joudef{\npb}    { Nuc.\ Phys.\ B}
\joudef{\ph}     { Physica}
\joudef{\pla}    { Phys.\ Lett. A}
\joudef{\plb}    { Phys.\ Lett. B}
\joudef{\pr}     { Phys.\ Rev.}
\joudef{\prd}    { Phys.\ Rev.\ D}
\joudef{\prep}   { Phys.\ Rep.}
\joudef{\prl}    { Phys.\ Rev.\ Lett.}
\joudef{\prsla}  { Proc.\ Roy.\ Soc.\ Lond.\ A}
\joudef{\ptp}    { Prog.\ Theor.\ Phys.}
\joudef{\ptps}   { Prog.\ Theor.\ Phys.\ Suppl.}
\joudef\rmp      { Rev.\ Mod.\ Phys.}
\joudef\spj      { Sov.\ Phys.\ JETP}

\catcode`@=11

\def\eqalign#1{\null\,\vcenter{\openup\jot\m@th
  \ialign{\strut\hfil$\displaystyle{##}$&$\displaystyle{{}##}$\hfil
      \crcr#1\crcr}}\,}
\def\meqalign#1{\null\,\vcenter{\openup\jot\m@th
  \ialign{\strut\hfil$\displaystyle{##}$&&$\displaystyle{{}##}$\hfil
      \crcr#1\crcr}}\,}
\catcode`@=12   

\newdimen\arrayruleHwidth
\makeatletter
\def\Hline{\noalign{\ifnum0=`}\fi\hrule \@height \arrayruleHwidth
  \futurelet \@tempa\@xhline}
\makeatother

\newcommand\thickbaselines{\baselineskip=20pt\lineskip=3pt\lineskiplimit=3pt}
\catcode`@=11
\def\cases#1{\left\{\,\vcenter{\thicknormalbaselines\m@th
             \ialign{$##\hfil$&\quad##\hfil\crcr#1\crcr}}\right.}
\def\matrix#1{\null\,\vcenter{\thickbaselines\m@th
    \ialign{\hfil$##$\hfil&&\quad\hfil$##$\hfil\crcr
      \mathstrut\crcr\noalign{\kern-\baselineskip}
      #1\crcr\mathstrut\crcr\noalign{\kern-\baselineskip}}}\,} 
\catcode`@=12   

\newcommand{\eprint}{\textsf} 

\newcommand\be{\begin{equation}} \newcommand\ee{\end{equation}} 
\newcommand\bd{\begin{displaymath}}\newcommand\ed{\end{displaymath}}

\newcommand{\e}{{\rm e}}

\newcommand\ts\textstyle

\def\undersim#1{\mathop{\vtop{\ialign{##\crcr
     $\hfil\displaystyle{#1}\hfil$\crcr\noalign
     {\kern1pt\nointerlineskip}\hbox{$\hfil\sim\hfil$}\crcr
     \noalign{\kern1pt}}}}}

\usepackage{amsmath,amssymb}

\newcommand{\ncd}{\newcommand} 
\ncd{\nms}{\negmedspace} 
\ncd{\nts}{\negthickspace} 
\ncd{\mcl}[1]{\mathcal{#1}} 
\ncd{\beq} {\begin{equation}} 
\ncd{\eeq} {\end{equation}} 
\ncd{\BE} {\begin{eqnarray}} 
\ncd{\EE} {\end{eqnarray}} 
\ncd{\rarr} {\rightarrow} 
\ncd{\larr} {\leftarrow} 
\ncd{\lrarr} {\leftrightarrow} 
\ncd{\lbeq}[1]  {\label{eq: #1}} 
\ncd{\refeq}[1] {(\ref{eq: #1})} 
\ncd{\mrm}    {\mathrm} 
\ncd{\nn}{\nonumber} 
\ncd{\mbf}[1] {{\mathbf #1}} 
\ncd\T{\frac{1}{2}h^{\mu\nu}p_\mu p_\nu} 
\ncd{\ms}{\mathstyle} 
\ncd{\ds}{\displaystyle} 
\ncd{\bmth}[1] {\mbox{\boldmath $#1$}} 
\ncd{\abs}[1] {|#1|} 
 
\newtoks\reportnoregister \newtoks\eprintnoregister
\newcommand{\reportnumber}[1]{\reportnoregister={#1}}
\newcommand{\eprintnumber}[1]{\eprintnoregister={#1}}

\reportnumber{\mbox{}} 
\eprintnumber{\mbox{}} 

\newcommand{\reportid}{
   \begin{minipage}{17cm}\vspace{-7.2cm}
     \begin{flushright}
      {\normalsize \the\reportnoregister \\[-.2cm]
            \eprint{\the\eprintnoregister}}\vspace{0.2cm}
     \end{flushright}
   \end{minipage}\hspace{-17cm} }

\catcode`@=11   
\def\title#1{\gdef\@title{\reportid#1}}
\catcode`@=12   

\begin{document} 

\reportnumber{USITP 2001-9}
\eprintnumber{gr-qc/0111066}

\title{Axial perturbations of general spherically symmetric spacetimes}
  \author{Max Karlovini\footnote{E-mail: \eprint{max@physto.se}} \\[10pt]
  {\small Department of Physics, Stockholm University}  \\
  {\small Box 6730, 113 85 Stockholm, Sweden} }
\date{}
\maketitle


\begin{abstract}{\normalsize
    The aim of this paper is to present a governing equation for first
    order axial metric perturbations of general, not necessarily
    static, spherically symmetric spacetimes. Under the
    non-restrictive assumption of axisymmetric perturbations, the
    governing equation is shown to be a two-dimensional wave equation
    where the wave function serves as a twist potential for the
    axisymmetry generating Killing vector. This wave equation can be
    written in a form which is formally a very simple generalization
    of the Regge-Wheeler equation governing the axial perturbations of
    a Schwarzschild black hole, but in general the equation is
    accompanied by a source term related to matter perturbations. The
    case of a viscous fluid is studied in particular detail.
    }\end{abstract}
\vspace{.5cm}
\centerline{\bigskip\noindent PACS: 4.30.Nk, 4.40.Dg, 4.40.Nr}

 
\section{Introduction}
To consider perturbations of various physical fields around a
spherically symmetric background is natural for several reasons. From
the mathematical point of view, the symmetries of the background
simplify the equations considerably and make them tractable for taking
the analytical treatment further and for carrying out the numerical
calculations needed without going far beyond standard methods. From
the physical point of view, many systems (in this paper, the ones in
mind are mainly astrophysical) owe their main structure to their
monopole moment, and may consequently be regarded as perturbations of
spherically symmetric systems to a good approximation. Moreover, as
far as e.g.\ gravitational and electromagnetic fields are concerned,
being non-radiative when restricted to be spherically symmetric, it is
of course highly interesting to see what types of radiation are
allowed for their more general neighbours.

The discussion of this paper will be restricted to integer spin
fields, satisfying linear differential equations defined on a
spherically symmetric background. All such fields are known to split
naturally into two parts that are not coupled by the equations of
motion, namely an odd parity (or axial) part and an even parity (or
polar) part. A standard way of separating out the angular dependence
from the equations of motion for such fields is to make use of scalar,
vector and tensor spherical harmonics (cf.\ 
\cite{zerilli:tensorharm}), which all can be defined in terms of the
scalar ones, i.e.\ $Y_l^m(\theta,\phi)$, and derivatives thereof. As
the final form of the equations are known not to depend on the
separation constant $m$, there is no mathematical restriction of
setting $m$ to zero from the outset, corresponding to only considering
axisymmetric fields. This was for instance the approach taken by
Chandrasekhar in his extensive treatment of gravitational
perturbations of Schwarzschild and Reisser-Nordstr\"om black holes
\cite{chandra:mtobh}, which summed up and added insight to the
original works of Regge and Wheeler \cite{rw:axial} and Zerilli
\cite{zerilli:polar}. Later, the same approach was taken by
Chandrasekhar and Ferrari \cite{cf:osc} when generalizing to static
spherically symmetric perfect fluid stellar models. The main focus of
this paper is to use the axisymmetric approach to present a general
treatment of the axial gravitational perturbations, valid for all
spherically symmetric backgrounds and with no restriction on the types
of matter perturbations. Only after this work was completed, this
author became aware of the work of Gerlach and Sengupta
\cite{gs:gaugeinv} who presented such a treatment for both polar and
axial perturbations, using a $2+2$ splitting scheme. However, new
approaches and results are given in the present paper. Firstly, the
method used to derive the final form of the perturbation equations,
being a rather straightforward application of the general covariant
form of the perturbed Einstein equations, differs from the one used in
\cite{gs:gaugeinv}. Here, the axial perturbation equations are first
covariantly reduced to a Maxwell-like equation on the background
geometry, valid not only for spherically symmetric backgrounds, but
for any background with a non-null and twist-free Killing vector that
is perturbed in such a way that the Killing vector obtains a twist to
first order. The Killing vector is in this work identified with the
axisymmetry generator of the neighbour to the spherically symmetric
background, but it could e.g.\ just as well be identified with the
time-like Killing vector of a stationary neighbour to a static
background, meaning that the Maxwell-like equation can also be used in
other contexts. Secondly, the further reduction of perturbation
equations is shown to always lead to a governing wave equation of the
Regge-Wheeler type $(-\partial_t^{\,2}+\partial_x^{\,2}-V)\psi = S$,
with $t$ and $x$ being harmonically conjugate coordinates on the
2-space orthogonal to the SO(3) group orbits, $V$ being a potential
only depending on the background geometry and $S$ being a source term
directly related to the perturbation of the stress-energy tensor. It
is found that the potential $V$ can be written in a form which is a
strikingly simple generalization of the form found by Chandrasekhar
and Ferrari for the case of static perfect fluids: the energy density
and pressure combination $\rho-p$ simply has to be replaced by
$\rho-p_\bot$, where $p_\bot$ is the radial pressure of a radially
moving observer who measures $\rho$ as the energy density. It is also
made clear that the wave function $\psi$ simply corresponds to a
two-dimensional twist potential of the axisymmetry generator, given of
course that the perturbations are axisymmetric. Thirdly, for static
perfect fluid backgrounds it is known that the source term $S$ has to
vanish unless the perturbation is stationary, in which case it
corresponds to adding a small rotation to the fluid ball. Since it
thus would be interesting to see what kinds of source terms are
allowed when considering more general fluids, we discuss in detail a
fluid with viscosity and find that $S$ is in general non-vanishing,
which corresponds to an interaction between shear and gravitational
oscillations.

In addition to the treatment of axial gravitational perturbations, we
also consider Maxwell and Klein-Gordon test fields on a general
spherically symmetric background. Also in these cases the equations of
motion are found to be reducible to Regge-Wheeler type wave equations
of the form given above. In fact, it is found that the potential $V$
for the three types of fields can be written in the common form $V=V_s
+ V_l$, where $V_s$ and $V_l$ depends quadratically on the spin $s$
and the multipole moment $l$, respectively. The $V_l$ part can notably
be identified with the centrifugal potential for null geodesics. Such
a splitting of the wave potential into a spin part and a centrifugal
(or angular momentum) part has earlier been discovered for
Scharzschild black holes \cite{ks:quasinormal}, but to the author's
knowledge it has not been shown to generalize to the general
spherically symmetric case.

\section{Spacetimes admitting a nearly twist-free Killing vector} 
If a spacetime $(M,g_{ab})$ admits a Killing vector $\eta^a$ which is
close to being twist-free, i.e.\ hypersurface orthogonal, it is
natural to treat it is as a perturbation of a spacetime admitting a
Killing vector which is \emph{exactly} hypersurface orthogonal. We
shall therefore consider a one-parameter family of
spacetimes with an associated family of Killing vectors $\eta^a$. The
metrics of these spacetimes will be written in the form
\begin{equation}\lbeq{aximetric}
  g_{ab} = \bot_{ab} + F\mu_a\mu_b
\end{equation}
where
\begin{equation} 
  F = \eta^a\eta_a \qquad\quad \mu_a = F^{-1}\eta_a
\end{equation}
and $\bot_a{}^b$ is the projection operator that projects onto
the tangent subspaces orthogonal to $\eta^a$. Before continuing with
deriving perturbation equations, it will be useful to write down some
formulas originating directly from the Killing vector equation
$\nabla_{(a}\eta_{b)} = 0$ in the case of $\eta^a$ being non-null
(i.e.\ $F\neq 0$). To begin with, we note that the two-form
$\nabla_{\!a}\eta_b$ can be decomposed as
\begin{equation}\lbeq{nablaetapre}
  \nabla_{\!a} \eta_b = \eta_{[a}v_{b]} + v_{ab} 
\end{equation}
where
\begin{equation}
  \eta^a v_a = 0 \qquad\quad \eta^a v_{ab} = 0 \qquad\quad v_{(ab)} = 0. 
\end{equation}
Contracting eq.\ \refeq{nablaetapre} by $\eta^b$ it follows that
\begin{equation}
  v_a = -\nabla_{\!a}\ln\abs{F}.
\end{equation}
In terms of $\mu_a = F^{-1}\eta_a$, eq.\ \refeq{nablaetapre} can be
equivalently expressed as
\begin{equation}
  \nabla_{\!a}\mu_b = \mu_{(a}v_{b)} + F^{-1}v_{ab}
\end{equation}
which immediately implies that $v_{ab}$ is proportional to the
exterior derivative of $\mu_a$;
\begin{equation}
  v_{ab} = \ts\frac12 F\Omega_{ab} \qquad\quad \Omega_{ab} = 2\nabla_{\![a}\mu_{b]}. 
\end{equation}
Hence, we conclude that $\nabla_{\!a}\eta_b$ is completely determined
by $F$, $\nabla_{\!a}F$ and the exact two-form $\Omega_{ab}$ as
\begin{equation}\lbeq{nablaeta}
  \nabla_{\!a}\eta_b = -\eta_{[a}\nabla_{\!b]}\ln\abs{F} +
  \ts\frac12 F\Omega_{ab}. 
\end{equation}
It should be noted that $F$ and $\mu_a$, and hence their exterior
derivatives $\nabla_{\!a}F$ and $\Omega_{ab}$, are Lie dragged by
$\eta^a$. Indeed,
\begin{equation}
  \mathcal{L}_\eta F = \eta^a\nabla_{\!a}(\eta^b\eta_b) =
  2\eta^a\eta^b\nabla_{\!(a}\eta_{b)} = 0 
  \qquad\quad \mathcal{L}_\eta \mu_a = F^{-1}\mathcal{L}_\eta\eta_a = 0 
\end{equation}
where $\mathcal{L}_\eta\eta_a = 0$ of course follows from
$\mathcal{L}_\eta\eta^a=0$ and $\mathcal{L}g^{ab} = 0$. Also, it
clearly follows from eq.\ \refeq{aximetric} that the projection
operator $\bot_a{}^b$ (with any index positioning) is also annihilated
by $\mathcal{L}_\eta$. Now, it is easily realized that the Killing
vector $\eta^a$ is hypersurface orthogonal (aligned with a gradient)
precisely when the two-form $\Omega_{ab}$ vanishes. Indeed, the
failure of hypersurface orthogonality can in four dimensions be
described by the twist vector \cite{ksmh:exact} (sometimes defined
with opposite sign as in \cite{wald:gr})
\begin{equation}
  \Omega^a = -\epsilon^{abcd}\eta_b\nabla_{\!c}\eta_d 
\end{equation}
which according to eq.\ \refeq{nablaeta} is dually related to
$\Omega_{ab}$ as
\begin{equation}\lbeq{Omegadual}
  \Omega^a = -\ts\frac12 F\epsilon^{abcd}\eta_b\Omega_{cd} 
  \qquad\quad \Omega_{ab} = F^{-2}\epsilon_{abcd}\Omega^c\eta^d. 
\end{equation}
Consequently, $\Omega^a$ also satisfies $\mathcal{L}_\eta\Omega^a=0$.

Let us now consider a one-parameter family of axisymmetric solutions
to the Einstein equations, with the aim of linearizing these equations
around a particular solution for which $\Omega_{ab}=0$. Referring to
$\lambda$ as the parameter of the family, the notation $\delta =
d/d\lambda$ will be used for derivation of tensor fields. The first
order metric purturbation tensor $\gamma_{ab} = \delta g_{ab}$ turns
out to split naturally into two parts according to
\begin{equation}
\begin{split}\lbeq{gammasplit}
  \gamma_{ab} &= {}^{+}\gamma_{ab} + {}^{-}\gamma_{ab} \\
  {}^{+}\gamma_{ab} &= (\bot_a{}^c\bot_b{}^d +
  \mu_a\mu_b\eta^c\eta^d)\gamma_{cd} = \delta\bot_{ab} + (\delta F)\mu_a\mu_b \\
  {}^{-}\gamma_{ab} &= 2\eta^c\mu_{(a}\bot_{b)}{}^d\gamma_{cd} =
    2F\mu_{(a}\delta\mu_{b)} = 2\eta_{(a}\delta\mu_{b)}.
\end{split}
\end{equation}
As always, in the absence of a natural identification of spacetime
points along the family of spacetimes (i.e.\ a preferred one-parameter
family of diffeomorphisms), $\delta$ is only defined up to a gauge
transformation 
\begin{equation}\lbeq{gaugetransf}
  \delta \rarr \delta + \mathcal{L}_\zeta 
\end{equation}
given in terms of an arbitrary vector field $\zeta^a$. However, in the
present case it is natural to only consider the families of
diffeomorphisms whose associated tangent space mappings takes the
Killing vector $\eta^a$ into ``itself'', corresponding to the
condition
\begin{equation}\lbeq{deltaeta0}
  \delta\eta^a = 0
\end{equation}
which clearly implies that we only allow for gauge transformations
with vector fields restricted to satisfy
\begin{equation}
  \mathcal{L}_\zeta\eta^a = -\mathcal{L}_\eta\zeta^a = 0. 
\end{equation}
It then follows from $\eta^a\mu_a=1$, $\eta^a\bot_{ab}=0$ and
$g^{ac}g_{cb}=\delta^a{}_b$ that
\begin{align}
  &\eta^a\delta\mu_a = 0 \\
  &\eta^a\delta\bot_{ab} = 0 \\
  &g_{ac}\,g_{bd}\,\delta\bot^{cd} = -\delta\bot_{ab} - 2\eta_{(a}\delta\mu_{b)}
\end{align}
which will simplify the calculations to come considerably. To derive
what we shall refer to as the axial perturbation equations, we split
the Einstein equations in the form $Z_{ab} := R_{ab} -
\kappa(T_{ab}-\frac12 T^c{}_c\,g_{ab}) = 0$ into its ``components''
orthogonal to $\eta^a$ according to
\begin{align} \lbeq{Zpolar1}
  \bot^{ac}\bot^{bd} Z_{cd} &= 0 \\ \lbeq{Zpolar2} 
  \eta^a\eta^b Z_{ab} &= 0 \\ \lbeq{Zaxial}
  \bot^{ab}\eta^c Z_{bc} &= 0.
\end{align}
As we are now about to apply $\delta$ to these equations, we shall
need the general form of $\delta R_{ab}$, which reads \cite{wald:gr}
\begin{equation}
  \delta R_{ab} = -\ts\frac12\nabla_{\!a}\nabla_{\!b}\gamma^c{}_c -
  \frac12\square\gamma_{ab} + \nabla^c\nabla_{(a}\gamma_{b)c}
\end{equation}
where $\square = \nabla^a\nabla_{\!a}$. Now, it does not take a too
frustrating amount of algebra to realize that when evaluating eqs.\ 
\refeq{Zpolar1} - \refeq{Zaxial} at $\Omega_{ab}=0$, using the above
derived formulae, the ${}^{-}\gamma_{ab}$ part of the metric
perturbation tensor $\gamma_{ab}$ satisfies eqs.\ \refeq{Zpolar1} and
\refeq{Zpolar2} identically, while on the other hand
${}^{+}\gamma_{ab}$ satifies eq.\ \refeq{Zaxial} identically. Hence it
follows that the perturbation equations separate ${}^{+}\gamma_{ab}$
and ${}^{-}\gamma_{ab}$, implying that those parts of the metric
perturbation can be treated separately. We shall here only focus on
the ${}^{-}\gamma_{ab}$ part, later to be identified with the axial
(or odd parity) part when we specialize to the case of a sperically
symmetric background. We hence insert $\gamma_{ab} =
{}^{-}\gamma_{ab}$ into eq.\ \refeq{Zaxial}, evaluate at
$\Omega_{ab}=0$ to find the Maxwell-like equation
\begin{equation}\lbeq{Maxwell-like}
  \nabla_{\!b}(FQ^{ab}) = \kappa J^a
\end{equation}
where
\begin{align}
  &Q_{ab} = \delta\Omega_{ab} \\
  &J^a = 2\delta T^a \qquad\quad T^a = \bot^{ab}\eta^c T_{bc}. 
\end{align}
It should be noted that since $\Omega_{ab}$ and $T^a$ vanish on the
background - the former field by assumption, the latter by the
twist-free Killing vector $\eta^a$ being an eigenvector of the Ricci
tensor and hence also of the stress-energy tensor (cf.\
\cite{wald:gr}) - it directly follows from eq.\ \refeq{gaugetransf}
that the perturbation fields $Q_{ab}$ and $J^a$ are invariant under
the gauge transformation \refeq{gaugetransf}. This can be recognised
as a lemma by Stewart and Walker \cite{sw:gaugeinv}: the linear
perturbation of a quantity is (identification) gauge invariant if that
quantity vanishes when unperturbed. Moreover, using $\delta\eta^a = 0$
it can easily be shown that $Q_{ab}$ and $J^a$ share the properties of
their respective nonlinearized fields of being orthogonal to, as well
as Lie dragged by, $\eta^a$. The perturbation equation
\refeq{Maxwell-like} should be accompanied by these conditions, as
well as the condition of $Q_{ab}$ being closed, which follows from the
closedness of $\Omega_{ab}$ and the fact that $\delta$ commutes with
exterior differentiation.  Moreover, as an integrability condition of
eq.\ \refeq{Maxwell-like}, it follows that the linearized matter
current $J^a$ is conserved,
\begin{equation}\lbeq{Jconserved}
  \nabla_{\!a}J^a = 0. 
\end{equation}
In the next section we will in fact not be working with the
Maxwell-like equation \refeq{Maxwell-like} directly, but rather with
its dual equation
\begin{equation}\lbeq{Maxwell-like-dual}
  2\nabla_{\![a}Q_{b]} = -\kappa J_{ab} 
\end{equation}
where
\begin{equation}
  Q^a = \delta\Omega^a \qquad\quad J_{ab} = \epsilon_{abcd}J^c \eta^d.
\end{equation}
The linearized twist vector $Q^a$ as well as the dual matter field
$J_{ab}$ can easily be shown to also have the properties of being
orthogonal to and Lie dragged by $\eta^a$. It should be noted that the
duality relations \refeq{Omegadual} for $\Omega_{ab}$ and $\Omega^a$
holds also for their linearized versions $Q_{ab}$ and $Q^a$, which is
due to the fact that the perturbation equations are evaluated on a
background where $\Omega_{ab}$ and $\Omega^a$ vanishes. Moreover, the
above relation between $J_{ab}$ and $J^a$ can be inverted into
\begin{equation}\lbeq{Jduality}
  J^a = -{\textstyle\frac12} F^{-1}\epsilon^{abcd}\eta_b J_{cd}.
\end{equation}
Finally, in this dual formulation, the additional equations
$\nabla_{\![a}Q_{bc]} = 0$ and $\nabla_{\!a}J^a = 0$ are replaced by
\begin{align}\lbeq{preRW}
  &\nabla_{\!a}(F^{-2}Q^a) = 0 \\ \lbeq{Jclosed}
  &\nabla_{\![a}J_{bc]} = 0.
\end{align}

\section{Axial gravitational perturbations}
\label{sec:axialpert}
We shall now restrict the reduced perturbation equations derived in
the previous section to the case of a spherically symmetric background
perturbed into an axisymmetric neighbour. As mentioned in the
introduction, the Killing vector $\eta^a$ will thus be chosen as 
the generator of the axisymmetry, reducing simply to an SO(3)
generator for the background.  The metric of the background geometry
will be written in the standard form
\begin{equation}\lbeq{ssmetric}
  ds^2 = d\sigma^2 + r^2(d\theta^2 + \sin^2\!\theta\,d\phi^2)
\end{equation}
where $d\sigma^2$ is the induced Lorentzian two-metric on the space
orthogonal to the SO(3) group orbits and $r$ is the Schwarzschild
radius. For simplicity, we will arrange the coordinates so that the
Killing vector $\eta^a$ will coincide with
$(\partial/\partial\phi)^a$. Note that the scalar $F = \eta^a\eta_a$
on this background takes the form
\begin{equation}
  F = (r\sin\theta)^2.
\end{equation}
To reduce the perturbation equations to two-dimensional ones by
separating out the dependence of the remaining angular variable
$\theta$, we start off by noting that no restricion is implied by
assuming that the two-form $J_{ab}$, which by eq.\ \refeq{Jclosed} is
closed, can be given in terms of a potential $Y_a$ which is orthogonal
to the SO(3) orbits (i.e.\ whose $\theta$ component is vanishing).
Thus we set
\begin{equation}\lbeq{Jdualpot}
  J_{ab} = 2\nabla_{\![a}Y_{b]}.
\end{equation}
It then follows from eq.\ \refeq{Maxwell-like-dual} that the
linearized twist vector $Q^a$ can be written in terms of $Y_a$ and a
scalar $\Phi$ as
\begin{equation}\lbeq{Qsep}
  Q_a = \nabla_{\!a}\Phi - \kappa Y_a.
\end{equation}
Turning to eq.\ \refeq{preRW} which corresponds to the closedness
condition for $Q_{ab}$, the $\theta$ dependence can be separated out
by setting 
\begin{equation}\lbeq{thetasep}
  \Phi = C(\theta)r\psi \qquad\quad Y_a = C(\theta)X_a
\end{equation}
where the scalar $\psi$ and the one-form $X_a$ have no angular
dependence, i.e.\ they are Lie dragged by all SO(3) generators.
Inserting eq.\ \refeq{Qsep} into eq.\ \refeq{preRW} leads to a
separation into an ordinary differential equation for $C(\theta)$ as
well a two-dimensional wave equation for $\psi$ coupled to $X_a$. The
equation for $C(\theta)$ will be discussed further in the appendix.
Here we merely note that it is solved by setting
\begin{equation}
  C(\theta) = G_{l+2}^{-3/2}(\cos\theta)
\end{equation} 
with $G_{l+2}^{-3/2}(y)$ being the ultraspherical (or Gegenbauer)
polynomial. The wave equation can be covariantly written in
form
\begin{equation}\lbeq{RWcov}
  (\mathcal{D}_a\mathcal{D}^a - U)\psi = \kappa S 
\end{equation}
where $\mathcal{D}_a$ is the connection of the Lorentzian
two-metric $d\sigma^2$. 
The potential $U$ can be put in the formally simple form
\begin{equation}
  U = {\textstyle\frac12}\kappa\,\tau - \frac{6m}{r^3} + \frac{l(l+1)}{r^2}
\end{equation}
with $\tau$ invariantly defined (given the spherical symmetry)
by $\frac12\kappa\,\tau$ being the eigenvalue function of the Ricci tensor with
respect to the eigenvector $\eta^a$ (or any other SO(3) generator),
i.e.\ 
\begin{equation}
  R^a{}_b\eta^b = \textstyle\frac12\kappa\,\tau\,\eta^a. 
\end{equation} 
The reason for defining $\tau$ with the factor $\frac12\kappa$
inserted as above is that the background Einstein equations imply that
$\tau$ is simply minus the trace of the $2\times 2$ block of the
stress-energy tensor orthogonal to the SO(3) orbits, an interpretation
which has a clearer physical significance. The function $m$ is the
spherically symmetric mass function \cite{ms:msmass}, invariantly
defined in terms of the Schwarzschild radius $r$ according to
\begin{equation}
  \nabla_{\!a}r\,\nabla^a r = 1-\frac{2m}{r}. 
\end{equation}
The source term $S$ on the left hand side of eq.\ \refeq{RWcov} is
given by
\begin{equation}\lbeq{Sdef}
  S = r\mathcal{D}_a(r^{-2}X^a)
\end{equation}
and is thus directly related to the perturbation of the stress-energy
tensor. We will have more to say about the wave equation
\refeq{RWcov}, but this will be postponed to section
\ref{sec:intspin}.

Having found the general solution to the conserved current condition
$\nabla_{\!a}J^a = 0$ by introducing a gauge potential $Y_a$ for the
dual field $J_{ab}$ according to eq.\ \refeq{Jdualpot}, it is of
interest to transform this solution back to $J^a$ which is more
directly related to the matter perturbations. Using eqs.\ 
\refeq{Jduality}, \refeq{Jdualpot} and \refeq{thetasep}, it follows
that $J^a$ can be expressed as
\begin{equation}\lbeq{divJsol}
  J^a = (r^2\sin\theta)^{-1}\left[C'(\theta)\mathcal{J}^a -
  C(\theta)\mathcal{J}\,(\partial/\partial\theta)^a\right]
\end{equation}
where
\begin{equation}\lbeq{JX2d}
  \mathcal{J}^a = \epsilon^{ab}X_b \quad \Leftrightarrow \quad X_a = \epsilon_{ab}\mathcal{J}^b 
  \qquad\quad \mathcal{J} = \epsilon^{ab}\mathcal{D}_a X_b
\end{equation}
with $\epsilon_{ab}$ being the Levi-Civita volume form of the induced
two-metric. It directly follows that $\nabla_{\!a}J^a = 0$ can be
solved by putting $J^a$ in the form given by eq.\ \refeq{divJsol} and
requiring that $\mathcal{J}^a$ and $\mathcal{J}$ satisfy
\begin{equation}\lbeq{divJ2d}
  \mathcal{D}_a\mathcal{J}^a = \mathcal{J}
\end{equation}
as implied be eqs.\ \refeq{JX2d}. It should be stressed that since the
form $J^a$ may be restricted by the types of matter studied, it may
not be practical to solve the conserved current condition by
expressing $\mathcal{J}^a$ and $\mathcal{J}$ in terms some $X_a$
according to eqs.\ \refeq{JX2d}. Rather, the two-dimensional equation
\refeq{divJ2d} in general has to be considered as nontrivial equations
of motion for the perturbed matter fields. Hence it is appropriate to
note that the source term of the wave equation \refeq{RWcov} can be
expressed in terms of $\mathcal{J}^a$ rather than $X_a$ as
\begin{equation}
  S = r\epsilon^{ab}\mathcal{D}_a(r^{-2}\mathcal{J}_b).
\end{equation}

Let us now discuss Israel's junction conditions \cite{israel:junction}
of the derived perturbation equations. Returning first to the general,
not necessarily spherically symmetric case, we assume that we are
given a non-null hypersurface $\Sigma$ with unit normal $n^a$, $n^a
n_a = \epsilon = \pm 1$, which may be given an arbitrary smooth
continuation off of $\Sigma$. For practical purposes we need only
consider the case when $\eta^a$ is tangent to the hypersurface, i.e.
$\eta^a n_a = 0$. The fields to be matched across $\Sigma$ are the
first and second fundamental forms defined as
\begin{align}
  j_{ab} &= g_{ab} - \epsilon\,n_a n_b \\
  k_{ab} &= j_a{}^c j_b{}^d \nabla_{\!c}n_d.
\end{align}
By using a gauge transformation generated by a vector field $\zeta^a$
satisfying
\begin{equation}
  \eta^a\zeta_a = 0 \qquad\quad \mathcal{L}_\eta\zeta^a = 0
\end{equation}
we may set $\delta n_a$ to zero. Such a gauge transformation only
affects the polar part of the perturbations, since it leaves the axial
field $\delta\mu_a = \delta(F^{-1}\eta_a)$ invariant. Hence we are
able to make the gauge choice $\delta n_a = 0$ without making any
gauge fixing in the axial sector that we are considering. For
simplicity, we therefore set $\delta n_a = 0$ to find that
\begin{equation}
  \delta j_{ab} = \delta g_{ab} = \gamma_{ab}.
\end{equation}
The relevant projection for the axial perturbation junction condition
is 
\begin{equation}
  \bot_a{}^b\eta^c\delta j_{bc} = F\delta\mu_a
\end{equation}
from which we conclude that all components of $\delta\mu_a$ should be
matched. Note, however, that $\delta\mu_a$ is not a gauge invariant
field, since $\mu_a$ is not zero on the unperturbed
spacetime. Namely, under a
gauge transformation generated by a vector field $\zeta^a$ such that
\begin{equation}
  \zeta^a = f\eta^a \qquad\quad \eta^a\nabla_{\!a}f = 0
\end{equation}
one finds that $\delta\mu_a$ transforms as
\begin{equation}\lbeq{dmugauge}
  \delta\mu_a \rarr \delta\mu_a + \mathcal{L}_\zeta\mu_a = \delta\mu_a
  + f\mathcal{L}_\eta\mu_a + (\eta^b\mu_b)\nabla_{\!a}f = \delta\mu_a
  + \nabla_{\!a}f.
\end{equation}
As far as the second fundamental form is concerned, the relevant
quantity to consider and match across $\Sigma$ is
\begin{equation}
  \delta(\bot_a{}^b\eta^c k_{bc}) = \ts -\frac12 F Q_{ab}n{^b} =
  -\frac12 F^{-1}\epsilon_{abcd}n{^b}Q{^c}\eta{^d}.
\end{equation}
Let us now specify these matching conditions to the spherically
symmetric case. Excluding the nonradiative case $l=1$ which will not
be treated here, it is implied by eqs.\ \refeq{Qsep} - \refeq{Sdef}
that $\delta\mu_a$ and $Q_{ab}$ take the forms
\begin{align}
  \delta\mu_a &= -\frac{C'(\theta)}{(l+2)(l-1)r^2\sin^3\!\theta}\, 
\epsilon_a{}^b[\mathcal{D}_b(r\psi) - \kappa X_b] + \nabla_{\!a}f \\
  Q_{ab} &= \frac{\psi\,C'(\theta)}{r^3\sin^3\theta}\,\epsilon_{ab} +
  \frac{2C(\theta)}{r^2\sin^3\!\theta}\,(\nabla_{\![a}\theta)\epsilon_{b]}{}^c
  [\mathcal{D}_c(r\psi) - \kappa X_c]
\end{align}
where the term $\nabla_{\!a}f$ in the expression for $\delta\mu_a$
corresponds to the freedom to make the gauge transformation
\refeq{dmugauge}. Now, since $\delta\mu_a$ is to be continuous and
since $\nabla_{\!a}f = \mathcal{D}_a f + \partial
f/\partial\theta\,\nabla_{\!a}\theta$, it directly follows that
$\partial f/\partial\theta$ has to continuous. Furthermore, the
possibility of adding to $f$ a $\theta$-independent term with
discontinuous gradient is excluded by the fact that
$C'(\theta)/\sin^3\!\theta$ is not constant for $l\neq 1$.
Consequently, the matching of $\delta\mu_a$ leads to the condition
that $\epsilon_a{}^b[\mathcal{D}_b(r\psi) - \kappa X_b]$ and hence
$\mathcal{D}_a(r\psi) - \kappa X_a$ be matched. For all practical
purposes we can assume that $n^a$ is orthogonal to the SO(3) orbits,
implying that $n^a\nabla_{\!a}\theta = 0$ and hence
\begin{equation}
  Q_{ab}n^b =
  \frac{\psi\,C'(\theta)}{r^3\sin^3\!\theta}\,\epsilon_{ab}n^b +
  \frac{C(\theta)}{r^2\sin^3\!\theta}\,n^b\epsilon_b{}^c
  [\mathcal{D}_c(r\psi)-\kappa X_c]\nabla_{\!a}\theta. 
\end{equation}
It directly follows that the matching of $Q_{ab}n^b$ leads to the
further condition that the two-scalar $\psi$ itself should be matched.

\section{Klein-Gordon and Maxwell fields}\label{sec:intspin}
In this section, we shall consider Klein-Gordon and Maxwell test fields
on a general spherically symmetric background and, as was done for the
axial gravitational perturbations treated in the previous section,
reduce the equations of motion to two-dimensional wave equations. 

Starting with the Klein-Gordon case, the equations of motion is taken
to be
\begin{equation}
  \square\Phi = \varepsilon_0 j
\end{equation}
where an unspecified source scalar $j$ multiplied by a coupling
constant $\varepsilon_0$ has been added to the free massless
Klein-Gordon equation. Restricting to axisymmetric fields, $\Phi$ as
well as $j$ will be assumed to be Lie dragged by $\eta^a$, one of the
SO(3) generators. Referring to the form \refeq{ssmetric} for the
general spherically symmetric metric, the $\theta$ dependence can
straightforwardly be separated out according to
\begin{align}
  &\Phi = C(\theta)r^{-1}\psi \\
  &j = C(\theta)r^{-1}S
\end{align}
with $\psi$ and $S$ being SO(3) invariant scalars. Taking $C(\theta)$
to be the Legendre polynomial $P_l(\cos\theta)$, the Klein-Gordon
equation reduces to the covariant two-dimensional equation
\begin{equation}
  (\mathcal{D}_a\mathcal{D}^a - U)\psi = \varepsilon_0 S 
\end{equation}
which is formally identical to the eq.\ \refeq{RWcov}, but with the
first two terms in the potential $U$ having different coefficients;
\begin{equation}
  U = -{\textstyle\frac12}\kappa\,\tau + \frac{2m}{r^3} + \frac{l(l+1)}{r^2}. 
\end{equation}

As for Maxwells equations
\begin{align}
  &\nabla_{\!b}F^{ab} = \varepsilon_{\!1}j^a \\ 
  &\nabla_{\![a}F_{bc]} = 0 \\
  &\nabla_{\!a}j^a = 0
\end{align}
with the electromagnetic field strength $F_{ab}$ and the current $j^a$
assumed to be Lie dragged by $\eta^a$, the hypersurface orthogonality
of $\eta^a$ makes it natural to make the splitting
\begin{align}
  &F_{ab} = P_{ab} + 2\mu_{[a}S_{b]} \\
  &j_a = J_a + J\mu_a
\end{align}
where all introduced fields are assumed to be orthogonal to $\eta^a$
as well as annihilated by $\mathcal{L}_\eta$. Indeed, using
$\nabla_{\![a}\mu_{b]} = 0$ leads to a decoupling into the two sets of
equations
\begin{align}\lbeq{maxwell_axial1}
  &F\nabla_{\!a}(F^{-1}S^a) = \varepsilon_{\!1}J \\ \lbeq{maxwell_axial2}
  &\nabla_{\![a}S_{b]} = 0
\end{align}
and
\begin{align}\lbeq{maxwell_polar1} 
  &\nabla_{\!b}P^{ab} = \varepsilon_{\!1}J^a \\ \lbeq{maxwell_polar2}
  &\nabla_{\![a}P_{bc]} = 0 \\ \lbeq{maxwell_polar3}
  &\nabla_{\!a}J^a = 0
\end{align}
corresponding to the axial and polar sector, respectively. 

Starting with the simpler axial case, we solve eq.\
\refeq{maxwell_axial2} by setting
\begin{equation}
  S_a = \nabla_{\!a}\Phi. 
\end{equation}
The angular dependence can now be immediately be separated out, this
time according to
\begin{align}
  &\Phi = C(\theta)\psi \\
  &J = C(\theta)S
\end{align}
with $C(\theta)$ taken to be the ultraspherical polynomial
$G_{l+1}^{-1/2}(\cos\theta)$ and $\psi$ and $S$ having no angular
dependence. The resulting two-dimensional equation again takes the
form \refeq{RWcov}
\begin{equation}
  (\mathcal{D}_a\mathcal{D}^a - U)\psi = \varepsilon_{\!1}S
\end{equation}
but with the potential $U$ in this case only consisting of the
centrifugal term
\begin{equation}\lbeq{maxwellU}
  U = \frac{l(l+1)}{r^2}.
\end{equation}
Turning to the polar case, which clearly can be treated similarly to
the axial gravitational perturbations, we introduce the fields $P^a$
and $J_{ab}$ dual to $P_{ab}$ and $J^a$ according to
\begin{align}
  &P^a = \textstyle -\frac12\epsilon^{abcd}\eta_b P_{cd} \qquad\quad P_{ab}
  = F^{-1}\epsilon_{abcd}P^c\eta^d \\
  &J_{ab} = \epsilon_{abcd}J^c\eta^d \qquad\quad J^a = \textstyle -\frac12
  F^{-1}\epsilon^{abcd}\eta_b J_{cd}.
\end{align}
This turns eqs.\ \refeq{maxwell_polar1} - \refeq{maxwell_polar3} into
\begin{align} \lbeq{mpd1}
  &2\nabla_{\![a}P_{b]} = -\varepsilon_{\!1}J_{ab} \\ \lbeq{mpd2}
  &\nabla_{\!a}(F^{-1}P^a) = 0 \\ \lbeq{mpd3}
  &\nabla_{\![a}J_{bc]} = 0. 
\end{align}
We start off by solving the closedness condition for $J_{ab}$
according to 
\begin{equation}
  J_{ab} = 2\nabla_{\![a}Y_{b]}
\end{equation}
with the condition that the gauge potential $Y_a$ be orthogonal to the
SO(3) orbits. According to eq.\ \refeq{mpd1}, the field $P_a$ can be
written as
\begin{equation}
  P_a = \nabla_{\!a}\Phi - \varepsilon_{\!1}Y_a
\end{equation}
for some scalar potential $\Phi$. To reduce the remaing equation
\refeq{mpd2} into a two-dimensional one, we set
\begin{equation}
  \Phi = C(\theta)\psi \qquad\quad Y_a = C(\theta)X_a
\end{equation}
with $C(\theta) = G_{l+1}^{-1/2}(\cos\theta)$ and with $\psi$ and
$X_a$ being SO(3) invariant fields. Again, we obtain a two-dimensional
wave equation of the form \refeq{RWcov}. As for the axial Maxwell
case, the potential $U$ is given by eq.\ \refeq{maxwellU}, while the
source function $S$ has the form
\begin{equation}
  S = \mathcal{D}_a X^a = \epsilon^{ab}\mathcal{D}_a\mathcal{J}_b.
\end{equation}
The field $\mathcal{J}_a$ here refers to the form of $J^a$,
\begin{equation}
  J^a = (r^2\sin\theta)^{-1}\left[C'(\theta)\mathcal{J}^a -
  C(\theta)\mathcal{J}\,(\partial/\partial\theta)^a\right]
\end{equation}
where, as implied by the duality between $J^a$ and $J_{ab}$,
\begin{equation}
  \mathcal{J}^a = \epsilon^{ab}X_b \qquad\quad \mathcal{J} =
  \epsilon^{ab}\mathcal{D}_a X_b. 
\end{equation}
In analogy with the axial gravitational perturbations, the conserved
current condition $\nabla_{\!a}J^a$ is thus found to reduce to the
two-dimensional equation
\begin{equation}
  \mathcal{D}_a\mathcal{J}^a = \mathcal{J}.
\end{equation}

Having found that the linear equations of motion for Klein-Gordon
fields, Maxwell fields (axial as well as polar) and axial
gravitational perturbations are all governed by a two-dimensional wave
equation of the form
\begin{equation}\lbeq{RWspin}
  (\mathcal{D}^a\mathcal{D}_a - U)\psi = \varepsilon S
\end{equation}
it is interesting to note that the potential $U$ can be naturally
split into a spin dependent part $U_s$ and an angular momentum
dependent part $U_l$ as
\begin{align}
  &U = U_s + U_l \\
  &U_s = (s-1)\left[{\textstyle\frac12}\kappa\,\tau - (s+1)\frac{2m}{r^3}\right] \\
  &U_l = \frac{l(l+1)}{r^2}
\end{align}
with $s$ taking the values $0$, $1$ and $2$ for scalar,
electromagnetic and linearized gravity fields, respectively. Note also
that the value of the coupling constant $\varepsilon$ on the left hand
side of eq.\ \refeq{RWspin} depends on the spin $s$ as well;
$\varepsilon = \varepsilon_s$ with $\varepsilon_2 = \kappa$. In
geometrical units $2\varepsilon_{\!1} = \varepsilon_2 = 8\pi$. Now, to
connect more closely to previous works on various static backgrounds,
let us choose as coordinates two harmonically conjugate functions
\cite{wald:gr} $t$ and $x$ to make the two-metric $d\sigma^2$ take the
manifestly conformally flat form
\begin{equation}
  d\sigma^2 = e^{2\nu}(-dt^2 + dx^2). 
\end{equation}
This turns the wave equation \refeq{RWspin} into the form
\begin{equation}\lbeq{RWharm}
  \left(-\frac{\partial^2}{\partial t^2} + \frac{\partial^2}{\partial
  x^2} - V \right)\psi = \varepsilon H 
\end{equation}
where the potential $V$ and the source $H$ are simply given by
rescaling the invariant potential $U$ and source $S$ with the
conformal factor $e^{2\nu}$, i.e.\
\begin{equation}
  V = e^{2\nu}U \qquad\quad H = e^{2\nu}S
\end{equation} 
implying of course that $V$ also naturally splits according to $V =
V_s + V_l$. The coordinates $t$ and $x$ are clearly not uniquely
defined, since the freedom of choosing them is in
correspondance with the infinite dimensional conformal group in two
dimensions, but in the static case there is a natural particular
choice. Starting out from Schwarzschild coordinates
\begin{equation}
  d\sigma^2 = -e^{2\nu}dt^2 + \left(1-\frac{2m}{r}\right)^{\!\!-1}\!\!dr^2
\end{equation}
with $\nu$ and $m$ being functions of $r$ only, the harmonically
conjugate variables can preferably be taken to be the static time $t$
and the Regge-Wheeler (or tortoise) radial variable
\begin{equation}
  r_* = \int\!e^{-\nu}(1-2m/r)^{-1/2}\,dr
\end{equation}
which gives $d\sigma^2 = e^{2\nu}(-dt^2 + dr_*^2)$. In this case, we
may turn eq.\ \refeq{RWharm} into a one-dimensional Schr\"odinger
equation (in general with a source term) by separating out the time
dependence in the standard manner according to
\begin{equation}
  \psi = e^{i\omega t}Z(r_*) \qquad\quad H = e^{i\omega t}Y(r_*) 
\end{equation}
which results in 
\begin{equation}\lbeq{Schreq}
  \left( -\frac{d^2}{dr_*^2} + V - \omega^2 \right)Z = -\varepsilon Y.
\end{equation}
Clearly, besides the source $Y$, this is a standard one-dimensional
Schr\"odinger equation with the squared frequency $\omega^2$
corresponding to the energy. 

The splitting of the potential $V$ into a spin dependent and an
angular momentum dependent part has previously been noted in the case
of a Schwarzschild background (cf.\ \cite{ks:quasinormal}), for which
the matter term $\tau$ of course vanishes and the mass function $m$ is
a constant, but it does not appear to be have been known to generalize
to the case of a general spherically symmetric background in the
simple manner found here. Indeed, whether or not the background is
static, the only formal modification of the potential $V$ compared to
the axial gravitational wave potential found by Chandrasekhar and
Ferrari for static perfect fluid backgrounds \cite{cf:osc}, is the
simplest possible: the matter combination $\rho-p$ has to be replaced
by $\tau$, minus the trace of the $2\times 2$ block of the
stress-energy tensor that is SO(3) orbit orthogonal. Clearly, $\tau$
can be expressed as $\rho - p_\bot$, with $\rho$ and $p_\bot$ being
the energy density and radial pressure measured by an arbitrary
radially moving observer. It should be noted, however, that whereas
the combination $\rho-p_\bot$ is invariant under a change of such an
observer, the same does in general not hold true for the quantities
$\rho$ and $p_\bot$ themselves.

\section{An application to fluids with viscosity}
As a concrete and physically relevant example, we shall here consider
the general Regge-Wheeler type wave equation for axial gravitational
perturbations of spherically symmetric backgrounds, in the case when
the matter can be described as a fluid with viscosity in the standard
relativistic manner. The stress energy tensor for a viscous fluid is
\cite{mtw:gravitation}
\begin{equation}\lbeq{Tvisco}
  T_{ab} = \rho u_a u_b + (p-\zeta\Theta)h_{ab} - 2\eta\sigma_{ab}
\end{equation}
where the various fields involved are the fluid four-velocity $u^a$,
the energy density $\rho$, the pressure $p$, the coefficient of bulk
viscosity $\zeta$, the coefficient of dynamic (or shear) viscosity
$\eta$, the projection operator $h_{ab} = u_a u_b + g_{ab}$, the
expansion $\Theta = \nabla_{\!a}u^a$ and the shear tensor $\sigma_{ab}
= (\nabla_{\!c}u_{(a})h_{b)}{}^c - \frac13\Theta h_{ab}$.  For the
background solution, the four-velocity $u^a$ will be assumed to be
orthogonal to $\eta^a$, meaning that the matter is not rotating around
the symmetry axis. In particular, in the spherically symmetric case
the fluid will be assumed to be radially moving, thus automatically
satisfying $\eta^a u_a = 0$. We now note that the vector $T^a =
\bot^{ab}\eta^c T_{bc}$ takes the form
\begin{equation}
  T^a = (\rho+p-\zeta\Theta)\,\eta^c u_c\,\bot^{ab}u_b -2\eta\bot^{ab}\eta^c\sigma_{bc}.
\end{equation}
Applying the perturbation operator $\delta$ to this vector, we arrive
at the following form of the linearized matter current $J^a$:
\begin{equation}
\begin{split}
  \textstyle\frac12 J^a = \delta T^a =
  (\rho+p-\zeta\Theta-2\eta\sigma)\beta\,u^a + 
  \eta\left[FQ^{ab}u_b - Fh^{ab}\nabla_{\!b}(F^{-1}\beta) -
  \beta\dot{u}^a\right]
\end{split}
\end{equation}
where
\begin{align}
  &\beta = \eta^a\delta u_a \\
  &\sigma = F^{-1}\eta^a\eta^b\sigma_{ab}.
\end{align}
The popping up of the two-form $Q_{ab}$ in the expression for $J^a$
originates in the perturbation of the shear $\sigma_{ab}$, being
defined in terms of the covariant derivative of the four-velocity
$u_a$. More precisely, we have used
\begin{equation}
  \delta(\nabla_{\!a}u_b) = \nabla_{\!a}\delta u_b +
  u_c\,\delta\Gamma^c{}_{ab} \qquad\quad
  \delta(\bot^a{}_c\bot^d{}_b\Gamma^c{}_{de}\eta^e) = \ts-\frac12 F Q^a{}_b
\end{equation}
where $\Gamma^c{}_{ab}$ refers to the Christoffel symbols. As we now
specialise to the case of a spherically symmetric background, the
decomposition of $J^a$ in accordance with eq.\ \refeq{divJsol} is
achieved by decomposing the scalar $\eta^a\delta u_a$ according to
\begin{equation}
  \eta^a\delta u_a = \sin^{\!-1}\!\theta\,C'(\theta)\beta
\end{equation}
where $C(\theta)$, as in section \ref{sec:axialpert}, refers to the
Gegenbauer polynomial $G_{l+2}^{-3/2}(\cos\theta)$, while $\beta$ is a
scalar without angular dependence. The resulting expressions for the
two-dimensional fields $\mathcal{J}^a$ and $\mathcal{J}$ are
\begin{equation}\lbeq{Jvisc2d_1}
  \mathcal{J}^a = 2r^2(\rho+p-\zeta\Theta-2\eta\,\sigma)\,\beta\,u^a - 2\eta\,r\left[\psi +
  r^3n{^b}\mathcal{D}_b(r^{-2}\beta) + r\,\dot{u}\,\beta \right]n^a
\end{equation}
\begin{equation}\lbeq{Jvisc2d_2}
  \mathcal{J} = -2\eta\left\{n^a\mathcal{D}_a(r\psi) + \left[(l+2)(l-1)
  + 2\kappa\,r^2(\rho+p-\zeta\Theta-2\eta\,\sigma)\right]\beta\right\}
\end{equation}
where $n_a = u^b\epsilon_{ba}$, implying that $n^a$ is the (up to
sign) unique unit space-like vector which is orthogonal to both $u^a$
as well as the SO(3) orbits. We have also introduced the notation
$\dot{u}$ for the norm of the acceleration $\dot{u}^a$, and used that
$\dot{u}^a = \dot{u}\,n^a$. For simplicity, we shall now further restrict
the background by requiring that it be static. Using Regge-Wheeler
variables, the two-metric $d\sigma^2$ and the unit vectors $u^a$ and
$n^a$ take on the forms
\begin{align}
  &d\sigma^2 = e^{2\nu}(-dt^2+dr_*^2) \\
  &u^a = e^{-\nu}(\partial/\partial t)^a \\
  &n^a = e^{-\nu}(\partial/\partial r_*)^a
\end{align}
with the gravitational potential $\nu$ being a funcion of $r_*$
only. The time dependence of the perturbation fields is
straightforwardly separated out according to 
\begin{equation}
  \psi = e^{i\omega t}Z(r_*) \qquad\quad \beta = e^{i\omega t}B(r_*).
\end{equation}
The assumption of staticity implies that both the expansion $\Theta$
as well as the shear $\sigma_{ab}$ vanishes on the unperturbed
spacetime. This means that both $\Theta$ and $\sigma$ will be set to
zero in eqs.\ \refeq{Jvisc2d_1} and \refeq{Jvisc2d_2}, which inserted
into the matter current integrability condition \refeq{divJ2d} leads
to the ordinary differential equation
\begin{equation}\lbeq{divJ1dvisc}
  \left( -\frac{d^2}{dr_*^2} - e^\nu \dot{u}\frac{d}{dr_*} + W
  \right)B = e^{2\nu}\frac{\dot{u}}{r}\,Z
\end{equation}
where, after using the unperturbed Einstein equations, 
\begin{equation}
  W = e^{2\nu}\left[ \frac{\kappa}{2}(\rho+5p) + \frac{2m}{r^3} +
  \frac{l(l+1)}{r^2} - 2\sqrt{1-\frac{2m}{r}}\frac{\dot{u}}{r} \right]
  + \frac{i\omega}{\eta}\,\e^\nu(\rho+p).
\end{equation}
In obtaining eq.\ \refeq{divJ1dvisc}, we have multiplied eq.\
\refeq{divJ2d} by the coefficient of shear viscosity $\eta$. The case
$\eta = 0$ is hence recovered from eq.\ \refeq{divJ1dvisc} by first
multiplying by $\eta$ and thereafter setting $\eta=0$. This results in
the simple condition
\begin{equation}
  \omega B = 0
\end{equation}
meaning that for vanishing coefficient of shear viscosity, the
perturbation is either stationary ($\omega = 0$) or can be interpreted
as a pure gravitational wave not interacting with the matter ($B =
0$). This is in agreement with Chandrasekhar and Ferrari's work on
axial perturbations of isentropic perfect fluids \cite{cf:osc}, but
note that we here have not assumed that the coefficient of bulk
viscosity vanishes, so the fluid need not be an isentropic one for the
same result to hold. 

Returning to the case $\eta\neq 0$, the Schr\"odinger equation
\refeq{Schreq} (with $s=2$), the source function
$Y$ can be written as
\begin{equation}
  Y = -2r\frac{d}{dr_*}\left[(\rho+p)e^\nu B\right] + 2i\omega\,\eta\left[r^3
  e^{-\nu}\frac{d}{dr_*}(r^{-2}e^\nu B) + e^\nu Z\right]
\end{equation}
where it has been used that $\dot{u} = e^{-\nu}d\nu/d r_*$ to put $Y$
into a compact form. Since $Y$ contains the wave function $Z$, it is
clearly appropriate to rewrite eq.\ \refeq{Schreq} by moving the term in
question to the left hand side, with the result
\begin{equation}\lbeq{RWvisc}
  \left[ -\frac{d^2}{dr_*^2} + V -
  \omega(\omega-2i\kappa\,\eta\,e^\nu) \right]Z = -\kappa \tilde{Y}
\end{equation}
where
\begin{align}
  V &= e^{2\nu}\left[{\textstyle\frac12}\kappa\,(\rho-p) -
  \frac{6m}{r^3} + \frac{l(l+1)}{r^2}\right] \\
  \tilde{Y} &= -2r\frac{d}{dr_*}\left[(\rho+p)e^\nu B\right] +
  2i\omega\,\eta\,r^3
  e^{-\nu}\frac{d}{dr_*}(r^{-2}e^\nu B).
\end{align}
To summarize, the axial perturbations of a static sperically symmetric
viscous fluid are governed by the system of two coupled ordinary
differential equations \refeq{RWvisc} and \refeq{divJ1dvisc}. To solve
the equations numerically, one starts at the center, using as initial
conditions the following expansions in the Schwarzschild radius of the
solution with regular center:
\begin{align}
  Z &= r^{l+1}(Z_0 + Z_1 r^2 + \ldots\,) \\
  B &= r^{l+1}(B_0 + B_1 r^2 + \ldots\,) 
\end{align}
where
\begin{align}
  (2l+3)Z_1 &= \left\{\frac{\kappa}{4}(l+2)\left[ \frac13 (2l-1)\rho_0-p_0 \right] -
  \frac12\omega_0(\omega_0-2i\kappa\,\eta) \right\} Z_0 \\ \nn
  &- \kappa\left[\,(l+1)(\rho_0+p_0) - (l-1)i\omega_0\eta\,\right]B_0
  \\[8pt]
  (2l+3)B_1 &= \left\{
  \frac{\kappa}{4}\left[\left(1+\frac23l(l+1)\right)\rho_0 -
  (2l-1)p_0\right] + \frac{i}2\frac{\omega_0}{\eta}(\rho_0+p_0)
  \right\}B_0 \\ \nn
  &-\frac{\kappa}{12}(\rho_0+3p_0)Z_0 \\[8pt]
  \omega_0 &= e^{-\nu_0}\omega\,.
\end{align}
To arrive at these expansion coefficients we have used 
\begin{equation}
\begin{split}
  \rho &= \rho_0 + \rho_1 r^2 + \ldots \\
  p &= p_0 + p_1 r^2 + \ldots \\
  m &= \ts\frac16\kappa\rho_0 + m_1 r^2 + \ldots \\
  \nu &= \nu_0 + \ts\frac1{12}\kappa(\rho_0 + 3p_0)r^2 + \ldots
\end{split}
\end{equation}
which follows from the background Einstein equations. The integration
of the perturbation equations should be continued up to the stellar
surface at $r=R$, with $R$ denoting the Schwarzschild radius of the
star. While the function $B$ has no vacuum analogue, the integration
of the function $Z$ should be further continued into the vacuum region
$r>R$ where eq.\ \refeq{RWvisc} is replaced by the original
Regge-Wheeler equation
\begin{equation}
  \left[ -\frac{d^2}{dr_*^2} + \left(1-\frac{2M}{r}\right)\left(-\frac{6M}{r^3}+\frac{l(l+1)}{r^2}\right) - \omega^2 \right]Z = 0
\end{equation}
where $M = m|_{r=R}$. The general condition that $\psi$ and
$\mathcal{D}_a(r\psi)-\kappa X_a$ be matched across $r=R$ boils down to
\begin{align}
  &Z|_{r\rarr R^-} = Z|_{r\rarr R^+} \\
  &\left[\frac{d}{dr_*}(rZ) + 2\kappa e^\nu r^2(\rho+p)B\right]_{r\rarr
  R^-} = \left[\frac{d}{dr_*}(rZ)\right]_{r\rarr R^+}  \\
  &\left[\frac{d}{dr_*}(r^{-2}e^\nu B)\right]_{r\rarr R^-} = 0.
\end{align}
Furthermore, if the condition of purely outgoing gravitational waves
at infinity is imposed, corresponding to $Z\propto e^{-i\omega r_*}$
as $r\rarr\infty$, one is lead to a discrete gravitational wave
spectrum for the system, much in the same way as for perfect fluids
stars. The difference is that the gravitational waves (closely related
to the function $Z$) in this case interacts with fluid shear waves
(closely related to the function $B$) in the stellar interior.

\section{Concluding remarks}

So far, most of the works on gravitational perturbations of
spherically symmetric spacetimes have dealt with black holes or static
perfect fluid stellar models, with focus on the determination of the
quasinormal modes or means to excite them through various
astrophysical processes (cf.\ \cite{ks:quasinormal} for a review).
Gravitational radiation emitted by collapsing systems has also been
studied to some extent, notable examples being the work by Cunningham,
Price and Moncrief on Oppenheimer-Snyder colllapse
\cite{cpm:oddradiation, cpm:evenradiation} and by Seidel and Moore on
more general collapsing perfect fluids \cite{sm:collapsepert}. Since
Choptuik's discovery of critical phenomena in gravitational collapsing
systems \cite{choptuik:critical} and the subsequent development of the
subject into a brand new branch of general relativity, an increasing
number of different types of matter has been studied in the context
of, in particular, spherically symmetric gravitational collapse. Since
analysing perturbations of the critical solutions is a crucial part of
the game, it is clearly desirable to have a general framework for
perturbations of spacetimes with spherical symmetry, but without any
other restriction on neither the properties of the background nor on
the perturbations themselves. It is hoped that this work, being
complementary to Gerlach and Sengupta's \cite{gs:gaugeinv}, provides
such a framework. 

In this paper, we have put the axial gravitational perturbations on an
equal footing with Klein-Gordon and Maxwell test fields through the
unified two-dimensional wave equation presented in the previous
section. However, the more complicated polar gravitational
perturbations have not been discussed at all. Although some attempts
have been made to reduce the polar equations into a neat form, we have
so far not been able to take the equations any further than Gerlach
and Sengupta. However, judging from the vacuum case - a Schwarzchild
background perturbed into a more general vacuum neighbour - one may
suspect that a certain illuminating reformulation of the system of
two-dimensional polar equations may be possible in general. Indeed, as
discovered by Chandrasekhar \cite{chandra:mtobh}, the vacuum axial and
polar equations are related by what was later to be recognized as a
supersymmetry duality \cite{brink:gw}, more precisely meaning that the
Regge-Wheeler and Zerilli potentials for the axial respectively polar
modes are supersymmetric partners with the implication that it is only
necessary to investigate the properties of the former, simpler one.
This indicates that it could be possible that the polar equations for
the general spherically symmetric case can be reduced to an equation
formally identical to the generalized Regge-Wheeler equation found
here for the axial case, combined with a set of equations directly
related to the matter perturbations. Whether or not this is merely
wishful thinking remains an issue for further investigation.

We have looked into the case of viscous fluid stellar models in some
particular detail and derived a system of equations for the axial
perturbations that can be directly applied for numerical studies. It
would be interesting to explicitly calculate the w-modes for some
realistic models to see how the spectrum depends on the shear
viscosity $\eta$ which enters the equations. It would also be of
interest to do a similar analysis for other types of matter sources
that are more general than perfect fluids, such as elastic materials
as described in a general relativistic setting by Carter and Quintana
\cite{cq:elastica}.

\section*{Acknowledgements}
I am grateful to Professor Kjell Rosquist for helpful discussions.



\section*{Appendix: differential equation for the angular function $C(\theta)$}
For all integer spin fields studied in this paper, the separation of
the dependence on the angular variable $\theta$ leads to the ordinary
differential equation
\begin{equation}
  \left(\frac{d}{d\theta}\sin^{1-2s}\theta\frac{d}{d\theta} +
  k\sin^{1-2s}\theta\right)C(\theta) = 0
\end{equation}
where $s=0,1,2$ is the spin of the field and $k$ is a separation
constant. Setting $k = (l+s)(l-s+1)$, the general solution can be
written down as
\begin{equation}
  C(\theta) = (1-y^2)^{s/2}\left[C_1 P_l^s(\cos\theta) + C_2
  Q_l^s(\cos\theta)\right]
\end{equation}
where $P_l^s(y)$ and $Q_l^s(y)$ are the associated Legendre functions
of, respectively, first and second kind. To obtain solutions that are
regular at $y=\pm 1$, we must set $C_2$ to zero and $l$ to an integer
(which can be taken to be nonnegative since $(l+s)(l-s+1)$ is
invariant under $l\rarr -l-1$). The solution then becomes the
ultraspherical (or Gegenbauer) polynomial $G_{l+s}^{1/2-s}(y)$, which
for the radiative values of $l$, i.e.\ $l\geq s$, is related to the
Legendre polynomial $P_l(y)$ as
\begin{equation}
  G_{l+s}^{1/2-s}(y) = N_l^s(1-y^2)^s\frac{d^s}{dy^s}P_l(y)
\end{equation}
where $N_l^s$ is a normalization factor given by
\begin{equation}
\begin{split}
  &N_l^0 = 1 \\
  &N_l^s = \frac{\Pi_{j=1}^s(2j-1)}{\Pi_{j=1-s}^s(l+j)} \quad (s>0). 
\end{split}
\end{equation}
For the relevant values $s=0,1,2$, it may also be noted that the
Legendre polynomial can be expressed in terms of the ultraspherical
polynomial according to
\begin{equation}
\begin{split}
  &G_l^{1/2}(y) = P_l(y) \\
  &\frac{d}{dy}G^{-1/2}_{l+1}(y) = -P_l(y) \\
  &(l+2)(l-1)G_{l+2}^{-3/2}(y) + 2y\frac{d}{dy}G_{l+2}^{-3/2}(y) =
  -(1-y^2)P_l(y).  
\end{split}
\end{equation}

\end{document}